\documentclass[iop,numberedappendix,twocolappendix,appendixfloats]{emulateapjKA}
\usepackage{graphicx,subfigure,amsmath, amsfonts, amssymb,aas_macros,times}
\usepackage[usenames,dvipsnames,svgnames,table]{xcolor}
\usepackage[export]{adjustbox}
\usepackage[bookmarks=false]{hyperref}
\hypersetup{
  colorlinks= true, 
  urlcolor  = Blue, 
  filecolor=black,
  runcolor=blue,
  linkcolor = Red, 
  citecolor = RoyalBlue 
} 

\shorttitle{Mid-Infrared Properties of Post-starbursts}
\shortauthors{Alatalo et al.}
\slugcomment{Accepted to the Astrophysical Journal, May 11, 2017}

\begin{document}


\title{Welcome to the Twilight Zone: The Mid-Infrared Properties of Post-starburst Galaxies}

\author{Katherine Alatalo,$^{1}$\altaffilmark{$\dagger$} Theodoros Bitsakis,$^{2}$  Lauranne Lanz,$^{3,4}$ Mark Lacy,$^{5}$ Michael~J.\,I. Brown,$^{6,7}$\altaffilmark{$\ddagger$} K.~Decker French,$^{8}$ Laure Ciesla,$^{9}$ Philip~N. Appleton,$^{5}$ Rachael~L. Beaton,$^{1}$ Sabrina~L. Cales,$^{10}$ Jacob Crossett,$^{6}$ Jes\'us Falc\'on-Barroso,$^{11,12}$ Daniel~D. Kelson,$^{1}$ Lisa~J. Kewley,$^{13}$ Mariska Kriek,$^{14}$  Anne~M. Medling,$^{13,15}$\altaffilmark{$\dagger$} John~S. Mulchaey,$^{1}$ Kristina Nyland,$^{5}$ Jeffrey~A. Rich,$^{1}$ \& C.~Meg Urry$^{10}$}

\affil{
$^{1}$Observatories of the Carnegie Institution of Washington, 813 Santa Barbara Street, Pasadena, CA 91101, USA\\
$^{2}$Instituto de Radioastronom\'ia y Astrof\'isica, Universidad Nacional Aut\'onoma de M\'exico, C.P. 58190, Morelia, Mexico\\
$^{3}$Infrared Processing and Analysis Center, California Institute of Technology, Pasadena, California 91125, USA\\
$^{4}$National Radio Astronomy Observatory, 520 Edgemont Road, Charlottesville, VA 22903, USA\\
$^{5}$Dartmouth College, 6127 Wilder Laboratory, Hanover, NH 03755, USA\\
$^{6}$School of Physics, Monash University, Clayton, Victoria 3800, Australia\\
$^{7}$Monash Centre for Astrophysics, Monash University, Clayton, Victoria 3800, Australia\\
$^{8}$Steward Observatory, University of Arizona, 933 North Cherry Avenue, Tucson, AZ 85721, USA\\
$^{9}$Laboratoire AIM-Paris-Saclay, CEA/DSM/Irfu - CNRS - Universit\'e Paris Diderot, CEA-Saclay, F-91191 Gif-sur-Yvette, France\\
$^{10}$Yale Center for Astronomy and Astrophysics, Physics Department, Yale University, New Haven, CT 06511 USA\\
$^{11}$Instituto de Astrof\'isica de Canarias, E-38205 La Laguna, Tenerife, Spain\\
$^{12}$Departamento de Astrof\'isica, Universidad de La Laguna (ULL), E-38200 La Laguna, Tenerife, Spain\\
$^{13}$Research School of Astronomy and Astrophysics, Australian National University, Cotter Rd., Weston ACT 2611, Australia\\
$^{14}$Department of Astronomy, Campbell Hall, University of California, Berkeley, CA 94720, USA\\
$^{15}$California Institute of Technology, MC 249-17, 1200 East California Boulevard, Pasadena, CA 91125, USA
}
\altaffiltext{$\dagger$}{Hubble fellow}
\altaffiltext{$\ddagger$}{Future ARC fellow}
\email{kalatalo@carnegiescience.edu}

\begin{abstract}
We investigate the optical and {\sl Wide-field Survey Explorer} ({\em WISE}) colors of ``E+A'' identified post-starburst galaxies, including a deep analysis on 190 post-starbursts detected in the 2\micron\ All Sky Survey Extended Source Catalog. The post-starburst galaxies appear in both the optical green valley and the {\em WISE} Infrared Transition Zone (IRTZ). Furthermore, we find that post-starbursts occupy a distinct region  [3.4]--[4.6] vs. [4.6]--[12] {\em WISE} colors, enabling the identification of this class of transitioning galaxies through the use of broad-band photometric criteria alone. We have investigated possible causes for the {\em WISE} colors of post-starbursts by constructing a composite spectral energy distribution (SED), finding that mid-infrared (4--12\micron) properties of post-starbursts are consistent with either 11.3\micron\ polycyclic aromatic hydrocarbon emission, or Thermally Pulsating Asymptotic Giant Branch (TP-AGB) and post-AGB stars. The composite SED of extended post-starburst galaxies with 22\micron\ emission detected with signal to noise $\geq$3 requires a hot dust component to produce their observed rising mid-infrared SED between 12 and 22\micron. The composite SED of {\em WISE} 22\micron\ non-detections (S/N$<$3), created by stacking 22\micron\ images, is also flat, requiring a hot dust component. The most likely source of this mid-infrared emission of these E+A galaxies is a buried active galactic nucleus. The inferred upper limit to the Eddington ratios of post-starbursts are 10$^{-2}$--10$^{-4}$, with an average of 10$^{-3}$. This suggests that AGNs are not radiatively dominant in these systems. This could mean that including selections able to identify active galactic nuclei as part of a search for transitioning and post-starburst galaxies would create a more complete census of the transition pathways taken as a galaxy quenches its star formation.
\end{abstract}





\section{Introduction}
Galaxies in the modern universe show two bimodal distributions, one in morphology and one in color space. In morphology space, a ``tuning fork'' has been used to classify galaxies since \citet{hubble26}, representing spiral (late-type) galaxies and elliptical and lenticular (early-type) galaxies. In color space, galaxies break into a blue cloud and a red sequence \citep{baade58,holmberg58,tinsley78,strateva+01,baldry+04}, with a genuine dearth of galaxies with intermediate colors in the so-called ``green valley.'' This dearth is used to suggest that galaxies undergoing the metamorphosis between blue spirals and red early-types must be rapid.

Once {\em z}\,$\approx$\,0 galaxies begin the process of transitioning, the probability that it is a one-way process is high \citep{appleton+13,young+14}, with very few circumstances in which the galaxy will transition back permanently \citep{kannappan+13}. Because of this, it is essential to understand all possible pathways and physical mechanisms that can trigger a galaxy's metamorphosis. Many pathways to transformation have been observed, though it is likely that the list is not exhaustive. 

Mergers are capable of driving the molecular gas into the center, allowing it to be consumed in a starburst, and heating the stellar disks of the interacting galaxies \citep{toomre72,springel+05}, creating an elliptical galaxy. Minor mergers also appear capable of quenching star formation \citep{qu+10,eliche-moral+12,a14_stelpop}, especially if the recipient galaxy endures many minor mergers over its lifetime.  Secular evolution, in which a galaxy bulge grows sufficiently large to stabilize a molecular disk against gravitational collapse (thus inhibiting star formation) has also been shown in simulations to quench galaxies \citep{martig+09}, with additional observational evidence manifesting in early-type galaxies \citep{martig+13,davis+14}.

When galaxies fall into a cluster potential, they suffer strangulation, in which their ability to accrete external gas and replenish their supply is stunted \citep{bekki+02,blanton+moustakas09}, truncating star formation. They can also suffer harassment, in which gravitational torques from other cluster members dynamically heat the stars \citep{mihos95,moore+96,bekki98}. Group interactions \citep{hickson+92,zabludoff+mulchaey98} are able to catalyze quenching, which has been observed through the study of the individual group galaxies \citep{johnson+07,bitsakis+11,bitsakis+14,bitsakis+16,martinez-badenes+12,lisenfeld+14,a15_hcgco} and the evolution of the intragroup medium \citep{verdes-montenegro+01,rasmussen+08,borthakur+10}. Additionally, it is possible that much of the galaxy transformation observed in the cluster environment takes place during a group pre-processing phase \citep{dressler+13}.

\begin{figure}[t]
\raggedleft
\includegraphics[width=0.48\textwidth]{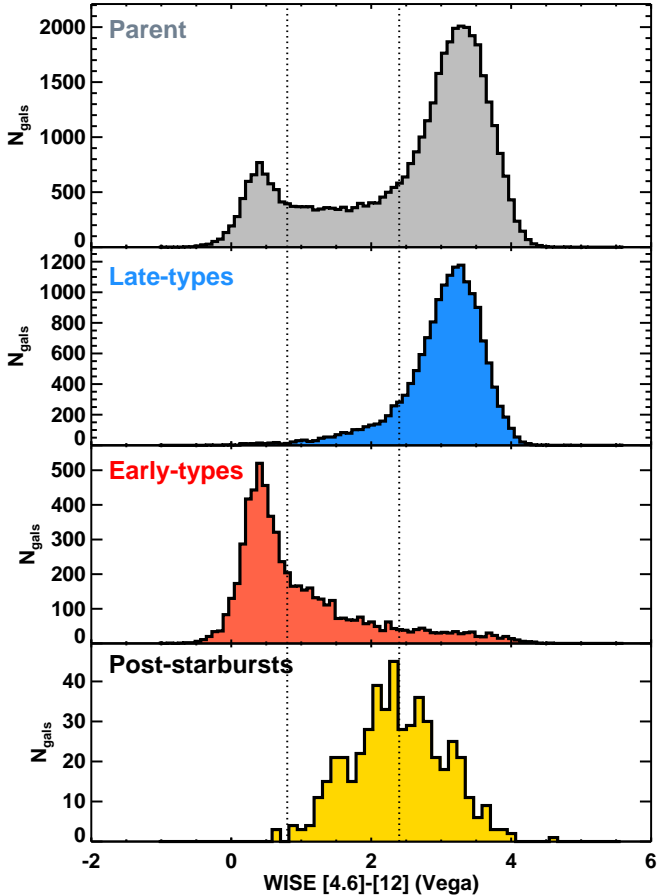} 
\caption{The [4.6]--[12] {\em WISE} color distributions of the various samples, including the Galaxy Zoo comparison sample of early-type and late-type galaxies (\citealt{schawinski+14,a14_irtz}; gray, top) delineated between late-type galaxies (blue) and early-type galaxies (red), with the redshift-corrected post-starburst galaxy colors for comparison (yellow).  The boundaries of the IRTZ are shown as a striped line through all plots. Post-starburst galaxies peak on the star-forming side of the IRTZ, and have the highest fractional IRTZ representation of the galaxies shown.}
\label{fig:psb_distribution}
\end{figure}

Active Galactic Nucleus (AGN) feedback, introduced to explain the truncated mass function of galaxies \citep{silk+98,dimatteo+05,croton+06,oppenheimer+10} can rapidly expel star-forming fuel from the galaxy and quickly quench star formation \citep{hopkins+06,hopkins+08}. Molecular gas outflows detected in some AGN hosts may be a signature of AGN feedback \citep{fischer+10,feruglio+10,sturm+11,alatalo+11,aalto_1377,aalto+16,cicone+12,cicone+14}, though the nearby examples do not appear to be powerful enough to rapidly eject the interstellar medium from the host, instead mainly injecting turbulence into the existing gas \citep{a15_sfsupp,guillard+15,lanz+15,lanz+16,costagliola+16} and ultimately depleting molecular gas at a rate consistent with the star formation rate \citep{alatalo15}. It is possible that radiation-mode AGN feedback provides sufficient energy to quench a galaxy at high redshift \citep{zakamska+16}, but the mechanism does not appear to be common in the modern universe. It is likely that the pathways discussed above are not an exhaustive sample; therefore creating a large sample of galaxies undergoing this transformation is necessary to probe the various conditions that can trigger it, possibly identifying new pathways that lead a galaxy to evolve.

\begin{figure*}[t]
\centering
\subfigure{\includegraphics[width=0.48\textwidth]{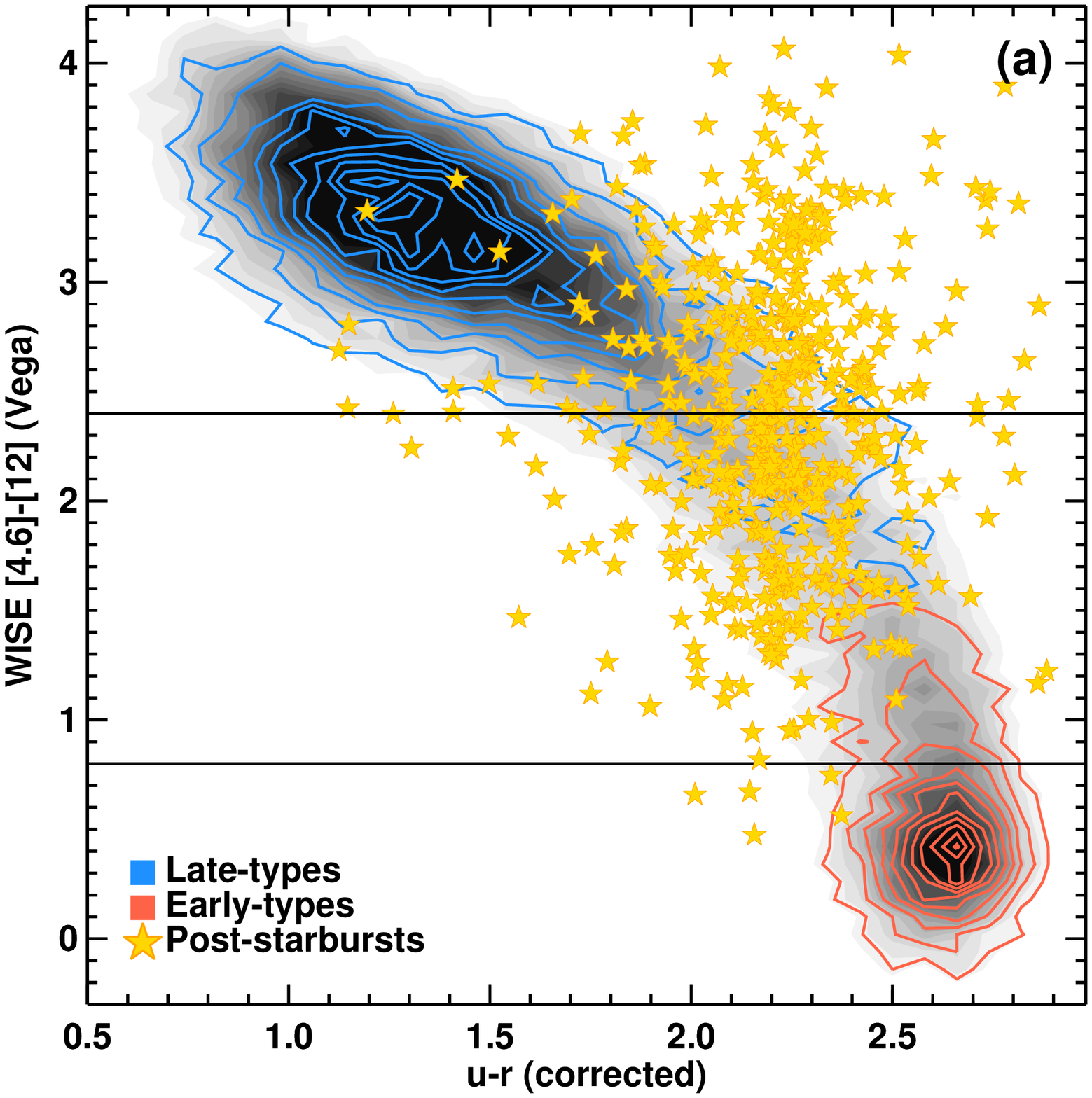}}
\subfigure{\includegraphics[width=0.48\textwidth]{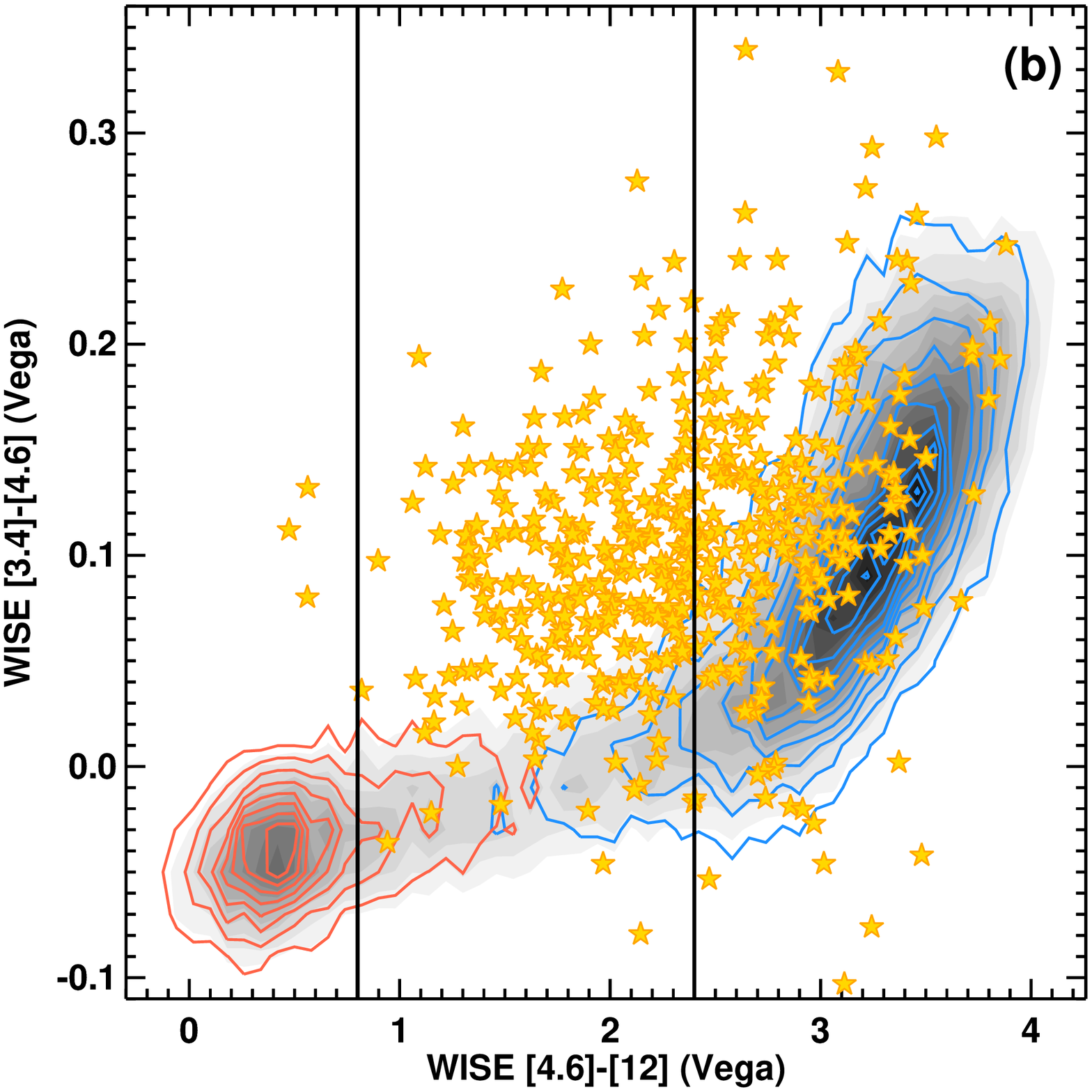}}\vskip -3mm
\caption{{\bf(Left):} The {\em u--r} vs. redshift-corrected [4.6]--[12] colors show a strong correlation with the prominent separation between late-types (blue contours) and early-types (red contours) from \citet{schawinski+14}. Post-starbursts (yellow stars) are located right between the two populations. {\bf(Right):} [4.6]--[12] vs. [3.4]--[4.6] {\em WISE} colors, which show that the post-starburst galaxies have slightly elevated [3.4]--[4.6] colors compared to the Galaxy Zoo sample. Post-starbursts have colors that are consistent with the Seyfert sample shown in \citet{a14_irtz}. The IRTZ is shown as black solid lines in each plot. The optical colors are {\em k}-corrected. The mid-IR colors are redshift corrected (see \S\ref{sec:zcorr}).}
\label{fig:psb_colors}
\end{figure*}

Despite the color and morphology bimodalities, finding galaxies that are rapidly transitioning is more complicated than determining their colors and morphologies. \citet{schawinski+14} showed that the number of galaxies within the green valley undergoing morphological change is small compared to galaxies whose intermediate colors are caused by secular processes, in which normal spiral galaxies with normal star-forming histories build up a substantial population of lower mass (redder) stars with a constant star formation rate, gradually turning the integrated colors of the galaxy green. 

More recently, a mid-infrared (mid-IR) color bimodality was observed using the {\sl Wide-field Infrared Survey Explorer} ({\em WISE}; \citealt{wise}). Authors identified a bimodality in both the [3.4]--[12]\micron\ \citep{ko+13,ko+16} as well as the [4.6]--[12]\micron\ colors \citep{yesuf+14,a14_irtz}. In the case of the [4.6]--[12] colors, \citet{a14_irtz} showed that color bimodality not only split based on galaxy morphology but that it was also more prominent than optical colors, and termed it the ``infrared transition zone'' (IRTZ).

Post-starburst galaxies are one such sample that have robustly been shown to have undergone a rapid cessation of star formation \citep{dressler+gunn83,zabludoff+96} via the presence of stellar absorption features consistent with intermediate stellar populations (such as strong Balmer absorption; \citealt{vazdekis+10}) and a lack of nebular ionized gas emission, such as H$\alpha$ or [O\,{\sc ii}]$\lambda$3727, which originates from H\,{\sc ii} regions associated with current (within the last 10\,Myr) star formation. These methods include the ``K+A'' method, which uses a weighting of A-star and K-star stellar libraries to determine a young star fraction \citep{dressler+gunn83,quintero+04} or a ``E+A'' identification (an early-type galaxy with A-type stars), which relies on Balmer absorption identification \citep{goto05,goto07}.  Although it is likely that the stringent selections used to pinpoint post-starburst galaxies miss a non-negligible fraction of transitioning galaxies, including those that host quasars \citep{canalizo+00,canalizo+13,cales+11,cales+13,cales+15} or shocks \citep{davis+12,a14_stelpop,a14_irtz,a16_sample}, they are a bonafide sample of transitioning galaxies.

We utilize the post-starburst sample compiled by \citet{goto07} to probe various properties of transitioning galaxies, including whether the {\em WISE} colors and the IRTZ \citep{a14_irtz} are able to identify a galaxy as having recently undergone a transformation. Given that post-starburst identification relies on available spectroscopy, being able to use photometry alone to pinpoint transitioning galaxies has the potential to substantially increase the total number of galaxies identified as undergoing this metamorphosis.

The paper is presented as follows. In \S\ref{sec:results}, we describe our post-starburst sample selection and comparison sample. In \S\ref{sec:disc}, we describe the post-starburst {\em WISE} properties and interpret those results. In \S\ref{sec:summary}, we summarize our findings. The cosmological parameters $H_0 = 70~$km~s$^{-1}$, $\Omega_m = 0.3$ and $\Omega_\Lambda = 0.7$ \citep{wmap} are used throughout.

\section{Results and Analysis}
\label{sec:results}

\subsection{Sample Selection}
\label{sec:sample}
We used the post-starburst galaxy sample defined by \citet{goto07} of 564 galaxies from the Sloan Digital Sky Survey Data Release 5 (SDSS DR5; \citealt{sdssdr5}), selected using the ``E+A'' criterion of deep Balmer absorption (EW(H$\delta$)\,$>$\,5\AA) combined with weak nebular (EW\,H$\alpha$\,$<$\,3\AA, and EW\,[O\,{\sc ii}]\,$<$\,2.5\AA) emission.  These objects have redshifts ranging between 0.03--0.34.  We cross-matched this sample with the {\em WISE} catalog \citep{wise} and the SDSS Data Release 9 (DR9; \citealt{sdssdr9}), using {\sc topcat} \citep{topcat}. Of the original 564 post-starburst galaxies, 560 have robust (S/N\,$>$\,3) detections in the W1/3.4$\mu$m, W2/4.6$\mu$m and W3/12$\mu$m bands. 534 objects are detected robustly in {\em u,\,r,\,i} filters.

In most cases, we used the profile fit ({\tt w$\star$mpro}) value from the {\em WISE} All-sky catalog for the {\em WISE} colors. When objects were flagged as extended, we elected to use {\tt w$\star$gmag}, which is the value derived using the 2-Micron All-Sky Survey (2MASS; \citealt{2mass}) profile fit, for the same aperture. The {\em u--r} colors are {\em k}-corrected using the {\tt calc\_kcor} IDL routine \citep{calc_kcor}\footnote{\href{http://kcor.sai.msu.ru/}{http://kcor.sai.msu.ru/}}.

In order to get a robust sub-selection of objects with near-IR data for a complete spectral energy distribution (SED), we cross-matched the 564 post-starburst galaxies from \citet{goto07} with the 2MASS Extended Source Catalog (XSC; \citealt{2mass}), containing the extended source photometries of 1.7 million galaxies\footnote{\href{http://www.ipac.caltech.edu/2mass/releases/allsky/doc/sec2\_3.html}{http://www.ipac.caltech.edu/2mass/releases/allsky/doc/sec2\_3.html}}.  In doing so, we recovered extended source photometries for 190 post-starburst galaxies. Then we cross-matched with the full-photometry catalogs of these samples (containing both SDSS and {\em WISE} data). Of the 190 post-starbursts, 158 were detected in the {\em WISE} 3.4, 4.6 and 12\micron\ bands. Only 53 of the post-starbursts were detected in the {\em WISE} 22\micron\footnote{The updated filter response function of the {\em WISE} W4 band places the central wavelength closer to 23\micron\ \citep{brown+14}, but we use 22\micron\ for consistency.} band with S/N\,$>$\,3. For subsequent color plots, we use the 534 E+A galaxies robustly detected in the SDSS {\em u\,r\,i} bands and {\em WISE} W1\,W2\,W3 bands. For subsequent composite spectral energy distribution (SED) plots, we use the 190 XSC E+A galaxies.

For our comparison sample, we use the morphologically classified Galaxy Zoo \citep{lintott+08} objects from \citet{schawinski+14}. We also cross-matched these 47,995 Galaxy Zoo objects with the 2MASS XSC, resulting in 38,802 Galaxy Zoo matches. The Galaxy Zoo comparison sample was drawn from \citet{schawinski+14} and \citet{a14_irtz}, and a corresponding analysis of the derivation of {\em WISE} colors can be found therein. Figures~\ref{fig:psb_distribution} \& \ref{fig:psb_colors} show the optical and {\em WISE} color distributions of both Galaxy Zoo and the post-starbursts.

\subsection{Redshift dependence of the WISE colors}
\label{sec:zcorr}

\begin{figure}[t]
\centering
\includegraphics[width=0.48\textwidth]{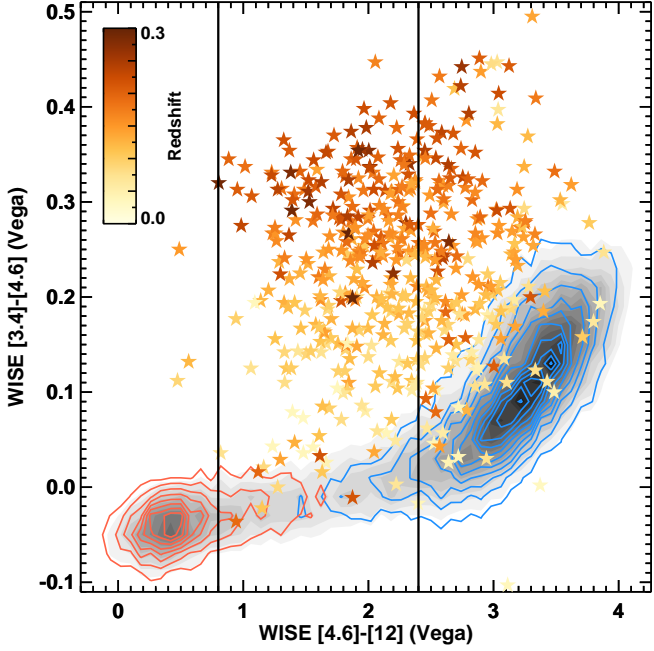} \vskip -1mm
\caption{The [4.6]--[12] vs. [3.4]--[4.6] {\em WISE} colors of post-starbursts (stars) as compared to the early-type (red contours) and late-type (blue contours) galaxies from \citet{schawinski+14,a14_irtz}. The redshift of the post-starbursts are encoded using the color scale on the upper left. A redshift dependence of the [3.4]--[4.6] colors is clear, with the most elevated colors coming from the highest redshift sources.}
\label{fig:w1w2_z}
\end{figure}

\begin{figure}[t]
\centering
\subfigure{\includegraphics[width=0.48\textwidth]{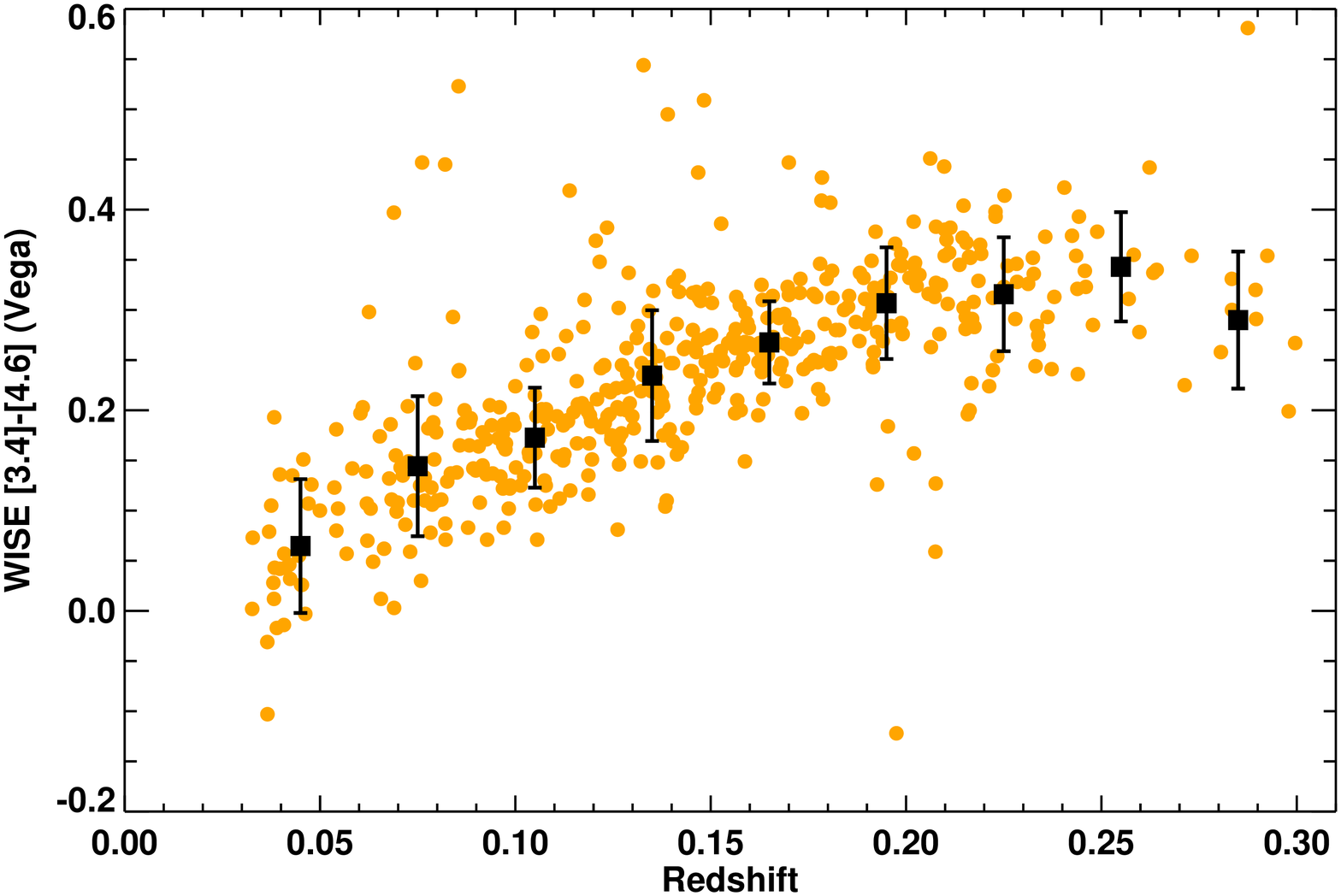}}
\subfigure{\includegraphics[width=0.48\textwidth]{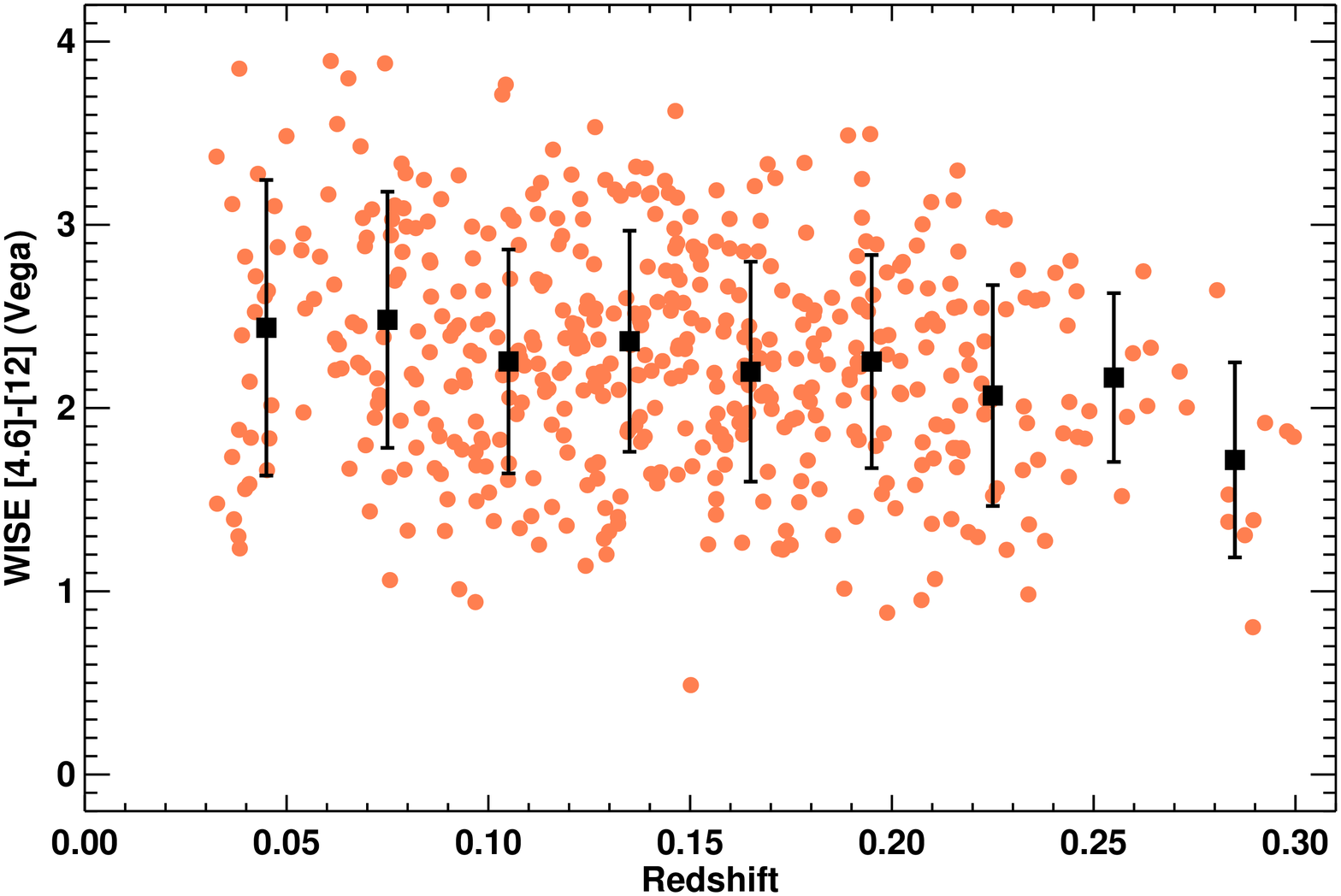}} \vskip -3mm
\caption{{\bf(Top):} The [3.4]--[4.6] colors of each of the 560 {\em WISE}-detected post-starbursts (orange points) versus redshift. The points are then binned in redshift, spaced in bins of {\em z}\,=\,0.03, ranging from {\em z}\,=\,0.03--0.3 with $\Delta${\em z}\,=\,0.03. The black squares represent the average [3.4]--[4.6] colors, with error bars representing their standard deviation. A clear redshift dependence is seen in these colors that is more substantial than the scatter in the bins.
{\bf(Bottom):} The [4.6]--[12] colors of each of the 560 {\em WISE}-detected post-starbursts (red points) versus redshift. Black squares represent the average colors in redshift bins with error bars representing the standard deviation. In this case, although there might be a slight trend with redshift, it is well within the (significant) scatter.}
\label{fig:z_dependence}
\end{figure}

Figure \ref{fig:w1w2_z} shows a significant deviation of the [3.4]--[4.6] {\em WISE} colors in post-starburst galaxies. Upon closer inspection, these colors have a substantial dependence on the redshift of the source, with the most significantly red colors having the highest redshifts. \citet{brown+14} showed that, given the SEDs of many galaxies go from decreasing to increasing in the mid-IR, accounting for a redshift dependence can be important. Figure~\ref{fig:z_dependence} shows the [3.4]--[4.6] and [4.6]--[12] colors for the post-starburst sample versus redshift, in both cases showing dependences. This is clear both in the individual post-starburst colors, as well as in the average colors in redshift bins. A slight redshift dependence is also seen in [4.6]--[12] colors, though the trend is smaller than the scatter in each bin. A significant redshift dependence in the [3.4]--[4.6] colors is seen, with a marked increase followed by a flattening at {\em z}\,$\geq$\,0.2.

There are many possible causes for this dependence including the effects of aperture bias, Malmquist bias \citep{malmquist25}, and and redshifting the SED. It is possible that the type of post-starburst galaxy that the \citep{goto07} criterion selected could have changed between the low redshift objects, where the SDSS fiber subtends a smaller fraction of the galaxy and the higher redshift objects, where much more of the galaxy is sampled. It is also possible that the \citet{goto07} selection detects brighter, rarer objects at higher redshifts. But given that the mid-IR is the location at which the SED transitions between the Rayleigh-Jeans tail of the stellar light of the galaxy and the hot dust component originating in the circumstellar envelopes of aging stars and shrouded star formation \citep{silva+98}, it is likely that this has the most dramatic effect on the [3.4]--[4.6] colors. A comprehensive SED fit to these galaxies is required to fully understand how each of these biases might impact our sample, and the constituents to the SED (discussed in \S\ref{sec:origin}) makes applying accurate {\em k}-corrections difficult.

To provide corrections to the [3.4]--[4.6] and [4.6]--[12] colors, we binned the post-starburst galaxies by redshift, every {\em z}\,=\,0.03, from 0.03 to 0.3 with $\Delta${\em z}\,=\,0.03. The average color of each bin was then determined for the colors, shown as the black squares in Fig.~\ref{fig:z_dependence}. The standard deviation of each redshift bin is reflected in the error bars. The Galaxy Zoo sample \citep{schawinski+14,a14_irtz} consists of objects with redshifts between {\em z}\,=\,0.02--0.05.  In order to reflect an accurate comparison to Galaxy Zoo, we ``correct'' the [3.4]--[4.6] and [4.6]--[12] colors of the post-starbursts by normalizing to the {\em z}\,=\,0.03--0.06 bin. We determine the redshift bin that each individual post-starburst galaxy sits in, then subtract the difference between the average color of that redshift bin from the average color of the {\em z}\,=\,0.03--0.06 bin. Figures~\ref{fig:psb_distribution} \& \ref{fig:psb_colors} use {\em WISE} colors that have been corrected this way.

\section{Discussion}
\label{sec:disc}
\subsection{The WISE colors of post-starburst galaxies}
\label{sec:colors}
Figure \ref{fig:psb_distribution} shows the distribution of the [4.6]--[12] colors of the Galaxy Zoo sample \citep{schawinski+14,a14_irtz} separated into the late-type and early-type subsamples with the {\em WISE} IRTZ overplotted.  As was shown in \citet{a14_irtz}, the Galaxy Zoo samples show a bimodal distribution, with a zone of avoidance between the late-type and early-type populations.  The post-starburst sample is strongly represented within the IRTZ, with 47.6$\pm$2.2\% falling within the bounds set in \citet{a14_irtz} (compared with 16.1$\pm$0.2\% of the Galaxy Zoo sample, 22.7$\pm$0.4\% of early-types and 10.5$\pm$0.2\% of the late-types), consistent with the hypothesis that the IRTZ is able to pinpoint galaxies that are transitioning.  The Mann-Whitney U test (IDL routine {\tt RS\_test}) was run to compare each pair of [4.6]--[12] {\em WISE} color distributions, which confirmed that post-starburst galaxies are a distinct population with a {\em p} value $\ll$\,10$^{-5}$ in all cases.

\begin{figure*}[t]
\centering
\includegraphics[width=0.95\textwidth]{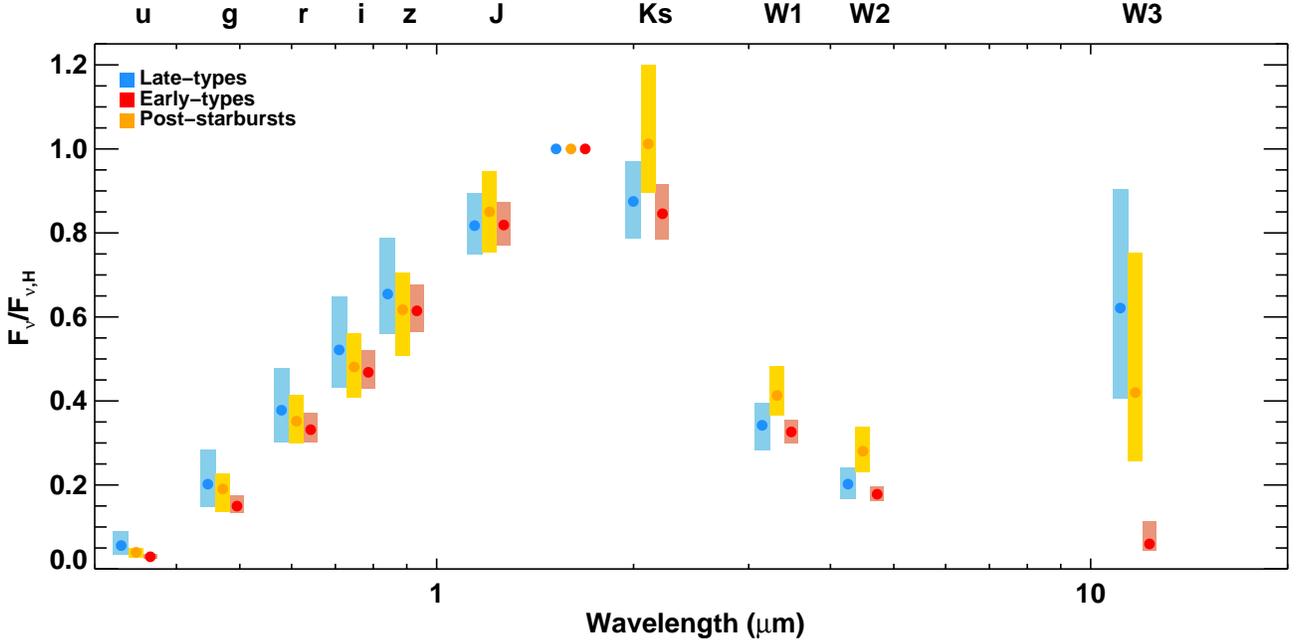} \vskip -1mm
\caption{The {\em u\,g\,r\,i\,z\,J\,H\,K$_s$} plus {\em WISE} band 1\,2\,3 {\bf observed frame} spectral energy distributions of each subsample: early-types (red), late-types (blue), and the 158 extended post-starbursts (yellow), normalized to their {\em H}-band fluxes. The width of the bar represents the upper and lower quartile, and the point is placed at the median value for each band. Post-starbursts contain colors which are in between early-type and late-type galaxies. Post-starbursts also show significant emission in 3.4 and 4.6\micron, with a shallower gradient. The 12$\mu$m flux in post-starbursts lies in between the late-type and early-type 12\micron\ fluxes, with much more overlap with the late-type galaxies.}
\label{fig:seds}
\end{figure*}

Figure \ref{fig:psb_colors} further supports this picture, placing post-starburst galaxies firmly within the transitioning region. Post-starbursts are located primarily in the optical green valley \citep{dressler+gunn83} and also appear in the {\em WISE} IRTZ.  The post-starburst population most obviously falls into the transition zone when viewed in {\em u--r} vs. {\em WISE} [4.6]--[12] color space, positioned amongst the tight color correlation between the early-type and late-type populations (Figure~\ref{fig:psb_colors}a). The post-starburst colors confirm that the {\em WISE} IRTZ traces a transitioning population and can be used as part of a criterion to identify galaxies through their photometry.

Post-starburst galaxies separate themselves into an elevated [3.4]--[4.6] vs. [4.6]--[12]$\mu$m {\em WISE} color space (Fig.~\ref{fig:psb_colors}b; also see Fig.~15 in \citealt{yesuf+14} for the colors of a similar sample). 48$\pm$2\% of galaxies fall outside of the 10\% contours of the Galaxy Zoo sample, and 91$\pm$1\% fall outside the 50\% contours of the Galaxy Zoo sample. These colors seem to indicate that post-starbursts do not traverse {\em WISE} color space through the joint between the early-type and late-types, rather showing signs of elevated [3.4]--[4.6] colors (and intermediate [4.6]--[12] colors) and traversing through a mid-IR ``twilight zone.'' The zone could be a consequence of the continued presence of ongoing star formation \citep{peletier+12,hayward+14}, or the presence of an AGN \citep{vanderwolk11,assef+13,mateos+15}.

\subsection{The origin of post-starburst WISE colors}
\label{sec:origin}
To further investigate the mid-IR twilight zone, we construct the average SEDs of the post-starbursts, early-types, and late-types. Figure~\ref{fig:seds} shows the SEDs of each of these subsamples, with the top and bottom of the bar representing the upper and lower quartiles, respectively. The point represents the median value for each band. Early-type galaxies are generally redder than late-type galaxies, and post-starbursts generally fall in between. In the {\em WISE} bands, the post-starbursts exhibit significantly stronger emission at [3.4] and [4.6] microns as well as a shallower gradient. The 12$\mu$m flux is generally in between that of the late-type and early-types, with much more overlap with the late-type galaxies.

\begin{figure*}[t]
\centering
\includegraphics[width=0.99\textwidth]{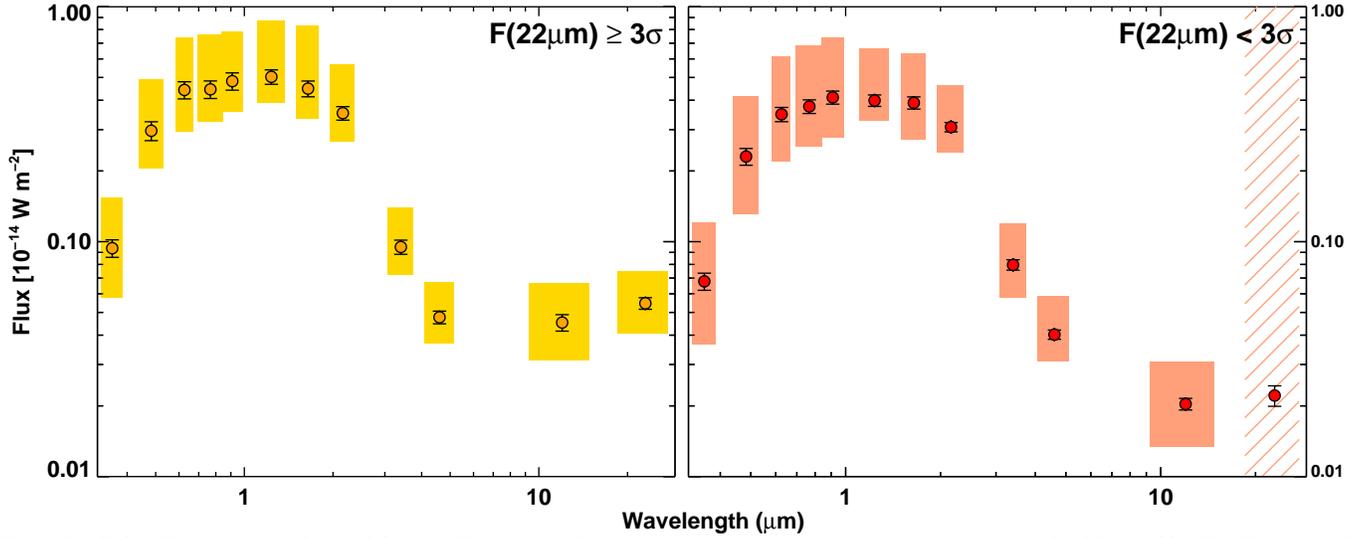}\vskip -2mm
\caption{{\bf(Left):} The composite {\bf observed frame} SED (yellow) of all post-starbursts in our extended catalog detected with S/N$>$3 ($N$\,=\,53). The width of the boxes represent the bandwidth of each filter, the upper and lower bounds of the boxes represent upper and lower quartiles of the distributions. The points represent the median, and the error bars represent the standard deviation of the median. {\bf(Right):} The composite SED (red) of all post-starbursts in our extended catalog detected with S/N$\leq$3 ($N$\,=\,105), with all features the same as the left panel, except the 22\micron\ point, which represents the value from the 4$\sigma$-clipped mean-stacked 22\micron\ emission. The most significant difference seen in the panels is the strength of the mid-IR (12 and 22\micron) emission, though in both cases, there is an observed flattening in the mid-IR SED.}
\label{fig:psb_seds}
\end{figure*}

Figure~\ref{fig:psb_seds} separates the SEDs of the post-starburst galaxies with (53) or without (108) detected 22$\mu$m emission. We created a composite SED (with the bars representing the upper and lower quartiles of the distribution, the point representing the median, and the error bars representing the standard deviation of the median) of the 22\micron\ detected objects, shown in the first panel of Figure~\ref{fig:psb_seds}. The 12\micron\ and 22\micron\ SEDs show a leveling off or increasing SED that cannot be reproduced using stars alone \citep{bruzual+charlot03}. While a compact starburst is capable of creating this rise in the SED, we consider this possibility unlikely. The compact starburst hypothesis assumes that the E+A selection is unable to remove these sources. Given that gas and extreme star formation coalesce into the nucleus of post-merged galaxies \citep{mihos+96,bryant+99}, it is unlikely that the SDSS spectrum would contain so little H$\alpha$ emission that it would make it through the E+A cut. Additionally, these buried compact starbursts are inconsistent with the 1.4\,GHz radio continuum results in K+A galaxies \citep{nielsen+12}.

The 22\micron\ emission in post-starbursts not individually detected significantly by {\em WISE} were stacked to create a composite image. Each 22\micron\ thumbnail was downloaded from the {\sc allwise}\footnote{\href{http://wise2.ipac.caltech.edu/docs/release/allwise/}{http://wise2.ipac.caltech.edu/docs/release/allwise/}} catalog \citep{wise}. The post-starburst source was placed in the center, and the stacked image was created by taking the average of stacked pixels (4$\sigma$ outliers in each pixel were clipped to remove contributions from bright stars in individual fields), resulting in a 22$\mu$m detection from the stack. To extract the photometry of the stacked 22\micron\ emission, we used an aperture with radius of 24$''$ (i.e., twice the {\em WISE} 22\micron\ resolution). We then subtracted the median emission from the ``sky'' in an annulus between 35$''$\,$<$\,$r$\,$<$\,55$''$. The resulting photometry (calculated and converted using the {\em WISE} manual\footnote{\href{http://wise2.ipac.caltech.edu/docs/release/allsky/expsup/sec4_4c.html}{http://wise2.ipac.caltech.edu/docs/release/allsky/expsup/sec4\_4c.html}}) is: $m$\,=\,15.86$^{+0.32}_{-0.25}$ in AB magnitudes. 

The photometry of each of the 190 XSC post-starburst galaxies was fit using {\sc magphys} \citep{magphys}, which is described in Appendix~\ref{app:magphys}. We isolate the {\sc magphys}-derived stellar and hot ($T_{\rm dust}$\,=\,130--250\,K) dust luminosity components for each post-starburst, and compare them to early-types, spirals, and AGNs.\footnote{Details on the comparison samples are discussed in Appendix~\ref{app:compsample}.} We find that [4.6]--[12] color from WISE correlates with and is therefore a good proxy for the ratio of the luminosities of the hot dust and stellar emission. This is somewhat expected, as the 4.6\micron\ emission primarily traces the stellar blackbody emission and the 12\micron\ emission primarily traces hot dust (from either star formation, aged stars, or AGNs).

\begin{figure*}[t]
\centering
\includegraphics[width=0.99\textwidth]{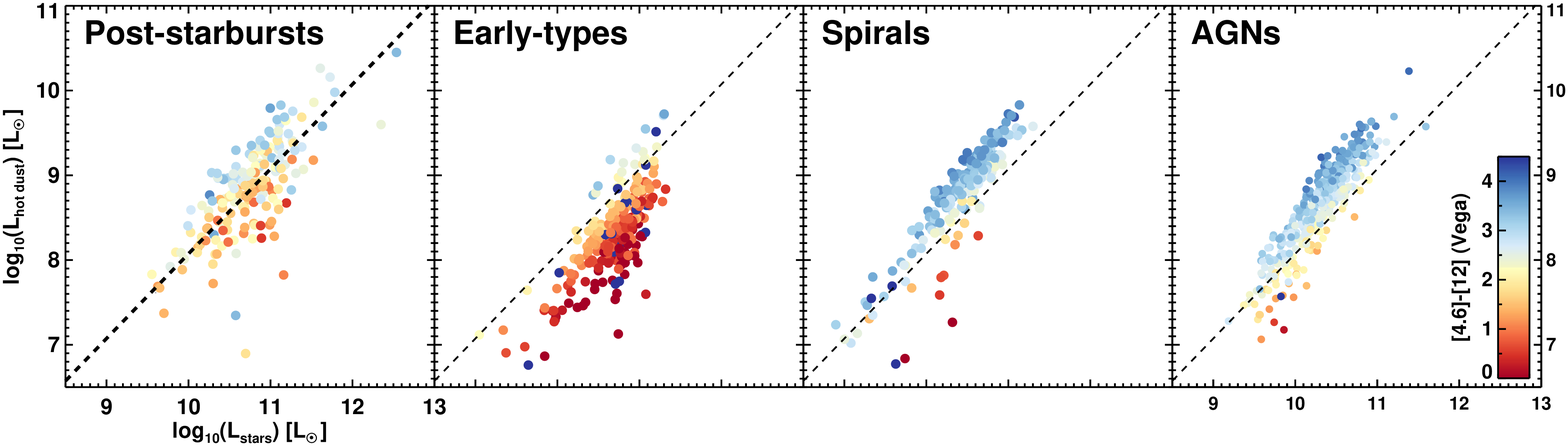} \vskip -1mm
\caption{The {\sc magphys}-derived stellar versus hot dust luminosity, color-coded based on the {\em WISE} [4.6]--[12]\micron\ colors of post-starburst (left), early-types galaxies (middle left), spirals (middle right), and AGNs (right). The stellar luminosity range is the same in all panels. The dashed black line corresponds to the mean relationship between the stellar and hot dust luminosities for the post-starbursts in all panels. The {\em WISE} [4.6]--[12]\micron\ colors are correlated with the relative strength of the hot dust (compared to stellar) luminosities. Early-type galaxies are the most stellar luminosity-dominant, and both spirals and AGNs have larger hot dust (compared to stellar) luminosities, though it is likely that the hot dust emission originates from a different source for each. Post-starbursts have hot dust-to-stellar luminosities that are between the early-type and spiral or AGN values.}
\label{fig:lum_comp}
\end{figure*}

Figure~\ref{fig:lum_comp} also shows that the post-starburst have middling luminosity ratios, between early-types and spirals/AGNs. Post-starbursts are expected to be in a transitional state between spirals and early-type galaxies (or possibly rejuvenated early-type galaxies; \citealt{dressler+13,abramson+14}), so perhaps their intermediate luminosities are unsurprising. This conclusion, however, assumes though that the hot dust emission has a similar origin to the hot dust emission in spirals (i.e., star formation). Given that the very selection of these galaxies preclude star formation from being significant, it is likely that the hot dust luminosity (and additionally the mid-IR emission from {\em WISE}) has an alternative heating mechanism. We lay out these possibilities in subsequent sections.

\subsubsection{PAHs in post-starbursts}
In normal star-forming galaxies, significant emission from photospheres and warm dust  and Polycyclic Aromatic Hydrocarbons (PAHs) in the interstellar medium \citep{calzetti+07,smith+07} usually overwhelms the mid-IR SED. Thus, in most star-forming galaxies, the {\em WISE} 12 and 22\micron\ bands are detecting active, recent star formation. In extreme starbursts (such as M82), there is a significant compact, hot dust component, leading to a rising SED  between 12 and 22\micron\ \citep{sturm+00,beirao+08}. Post-starbursts selected by \citet{goto07} removed all objects with significant H$\alpha$ emission; thus it is unsurprising that their SEDs do not match those of the star-forming galaxies in Fig.~\ref{fig:seds}, but the enhanced 12\micron\ emission (as compared to early-types) is worth discussing further.

By the time the galaxies have quenched and become completely quiescent, the total number of the evolved intermediate age stars has diminished. Therefore, the mid-IR emission in quiescent galaxies likely originates from a diffuse dust component \citep{temi+07,boselli+10,ciesla+14}, though usually the dust emission is dwarfed by the stellar emission. PAH emission is detectable in early-type galaxies \citep{xilouris+04,kaneda+05,kaneda+08,bressan+06,panuzzo+07,bregman+08} and is significant in K+A galaxies \citep{roseboom+09}. In both of these cases, the 12\micron\ emission is specifically enhanced due to the 11.3\micron\ neutral PAH feature, which could explain the significant 12\micron\ emission we see in Figure~\ref{fig:seds}. This was argued to be part of the existence of the IRTZ by \citet{a14_irtz}.

\citet{vega+10} presented the possibility that the unusual neutral-to-ionized PAH ratios that are observed are not due to accreted gas or an AGN but instead due to the processing of carbonaceous material from the circumstellar envelopes of Thermally Pulsating Asymptotic Giant Branch (TP-AGB) stars combined with slow shocks. These combination of these two processes may be able to create amorphous carbon and destroy the smaller PAHs (that create the shorter wavelength ionized PAH bands), leading to the enhanced 11.3\micron\ neutral emission. Thus, it is possible that 11.3\micron\ PAH emission (a potential source for the enhanced 12\micron\ emission) could originate from TP-AGB stars.

\subsubsection{AGB stars in post-starbursts}
\label{sec:agbs}
\begin{figure}[t]
\centering
\includegraphics[width=0.48\textwidth]{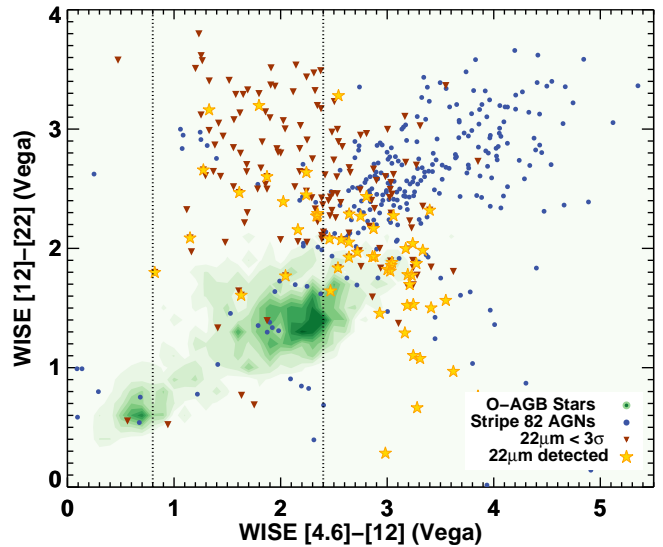}
\caption{The [4.6]--[12] vs. [12]--[22] {\em WISE} colors for Oxygen-rich AGB stars (green colorscale; \citealt{suh+11}), Stripe 82 ``strong'' AGNs (blue points; Glikman et al., in prep), 22$\mu$m non-detected (dark triangles) and 22$\mu$m-detected post-starbursts (yellow stars) from the extended source sample of 158. The IRTZ is represented by black dotted lines. As was shown in Figures~\ref{fig:psb_distribution} \& \ref{fig:psb_colors}, post-starbursts fall in the infrared transition zone. Many post-starbursts exhibit [12]--[22] colors consistent with AGNs.}
\label{fig:w2w3w4}
\end{figure}

Figure~\ref{fig:w2w3w4} presents the [4.6]--[12] vs. [12]--[22] {\em WISE} colors for Oxygen-rich asymptotic giant branch (AGB) stars \citep{suh+11}, Stripe 82\footnote{\href{http://classic.sdss.org/legacy/stripe82.html}{http://classic.sdss.org/legacy/stripe82.html}} ``strong'' AGNs\footnote{AGNs that are brighter than their hosts' starlight} (blue points; Glikman et al., in prep), 22$\mu$m non-detected (dark orange triangles), and 22$\mu$m-detected post-starbursts (yellow stars). For many of the post-starbursts (both 22\micron\ detected and non-detected), the [4.6]--[12] colors are consistent with oxygen-rich AGB stars, whose contributions to the optical spectra peak during the post-starburst phase of a galaxy \citep{yan+06}, and are inconsistent with the normal star-forming population (as is seen in the [4.6]--[12] color comparison in Fig.~\ref{fig:psb_colors}). 

Post-starburst galaxies are an ideal population to study mid-IR emission from the AGB population, given that their star formation has been quenched and no longer contributes, and the intermediate age stars are still abundant. Emission from circumstellar dust shells originating in TP-AGB stars tend to peak in the mid-IR \citep{piovan+03,maraston05,kelson+10,chisari+12}. A 2\,Gyr old AGB component leads to a slightly shallower slope in the mid-IR portion of the SED, which is observed in the post-starburst composite SED in Figure~\ref{fig:seds}. 

Previous studies of the TP-AGB and post-AGB contributions to post-starbursts have been contradictory, with optical spectra suggestive of their contribution \citep{yan+06} but IR spectra \citep{zibetti+13} and SED studies inconsistent with a dominant AGB component \citep{kriek+10,melnick+13}. In the cases of the contrasting studies, many used models that required a ``heavy'' AGB contribution that did not include circumstellar dust \citep{maraston05} and did not fit the SED out to the mid-IR. When mid-IR data are included, and the AGB model is modified to include circumstellar dust, TP-AGB and post-AGB models more consistently match the SEDs \citep{maraston+13}. The [4.6]--[12] colors from Fig.~\ref{fig:w2w3w4} seem to indicate that these stars could contribute to the mid-IR SED of post-starbursts, but the data are inconclusive. Thus, it is not clear what impact TP-AGB and post-AGB stars have on the integrated light of post-starburst galaxies; it will require further study to disentangle from other contributors to the 12\micron\ portion of a galaxy's SED.

\begin{figure}[t]
\raggedright
\includegraphics[width=0.48\textwidth]{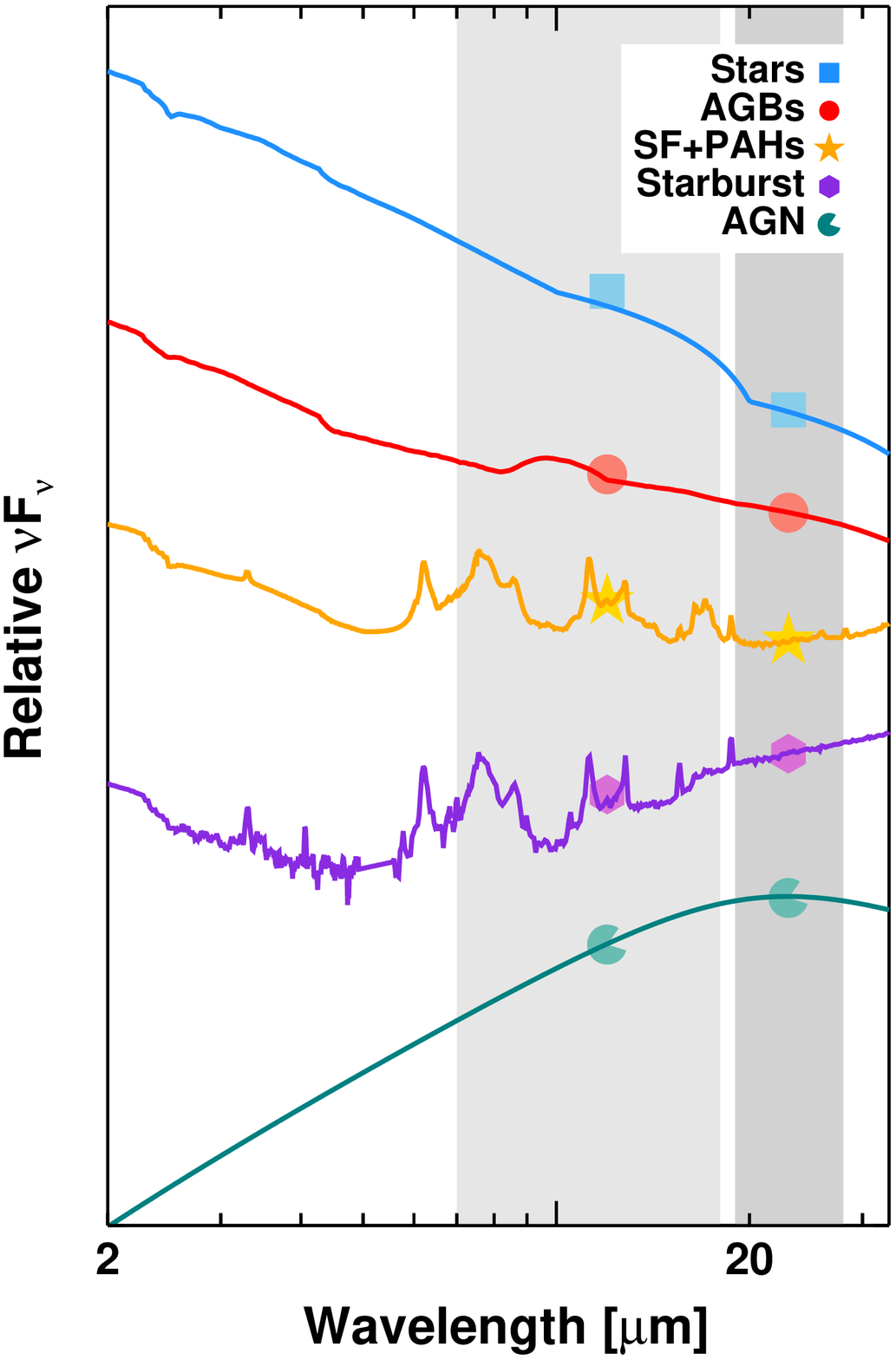} \vskip -3mm
\caption{The 2--30\micron\ region of various models of different objects, including a pure stellar population (blue; \citealt{bruzual+charlot03}), AGB stars (red; \citealt{piovan+03}), star-forming galaxies including PAH emission (orange; \citealt{brown+14}), starbursts (purple; \citealt{brown+14}), and an AGN (teal; \citealt{sajina+12}). The grayed regions represent the bandwidth of the {\em WISE} 12\micron\ and 22\micron\ filters. The blocks show the mean value of each model across the {\em WISE} bands. In all cases except for that of the starburst and AGN, it appears that the SEDs decrease with increasing wavelength.}
\label{fig:midIR_key}
\end{figure}

\subsubsection{AGNs in post-starbursts}
\label{sec:agn_psbs}
The {\em WISE} [3.4]--[4.6] vs. [4.6]--[12] colors of post-starburst galaxies shown in Figure~\ref{fig:psb_colors}b suggest that the transition in infrared color space of these objects is not a simple pathway in color space across the boundary that separates the early-type and late-type galaxy distributions. A robust exploration of the physical processes taking place in the quenching galaxies, and how those manifest in their integrated properties, can shed light on the way in which post-starburst galaxies undergo their metamorphosis. 

Post-starburst galaxies are thought to be the final stage of a transitioning galaxy. Evidence exists that there is a delay between star formation quenching and the onset of AGN activity \citep{canalizo+01,schawinski+07,kaviraj+15,matsuoka+15, bitsakis+16}; thus, we might expect post-starburst galaxies to disproportionatly host AGNs, assuming that they are just post-transition. The distribution of post-starburst {\em WISE} colors in fact show similarities to the Seyfert population discussed in \citet{a14_irtz}. Could this be a sign that the {\em WISE} colors of post-starburst galaxies originate from a buried AGN component?

Many studies into post-starbursts have aimed to confirm the presence of AGNs in these systems. \citet{brown+09} observed a slight enhancement in X-rays in a sample of K+A galaxies in the NOAO Deep Wide-field Survey, though not to a significance to be definitively from AGN emission. \citet{shin+11} cross-identify 1.4\,GHz FIRST \citep{first} sources with the \citet{goto07} E+A sample, but detections are ambiguous (most sources unresolved and lacking clear radio jets). \citet{nielsen+12} cross-referenced K+A galaxies with FIRST \citep{first}, and noted enhanced radio emission in some of them, which could be attributed to either AGN activity or remnant star formation. Though all of these studies could be pointing to low-luminosity AGNs being present, none are definitive. \citet{meusinger+17} studied a large sample of E+A galaxies, finding a slightly elevated number of luminous ($>$\,10$^{23}$\,W) 1.4\,GHz continuum sources and enhanced fraction of mid-IR {\em WISE} selected \citep{assef+13} AGNs, and confirmed the rising mid-IR SED discussed by \citet{melnick+13}, but concluded while E+A galaxies do not host strong AGNs, they may contain obscured and/or low-luminosity AGNs or require significant emission from post-AGB stars (as discussed in \S\ref{sec:agbs}).

The right panel of Figure~\ref{fig:psb_seds} shows the composite SED created from the 105 {\em WISE} 22\micron\ non-detected objects, with quartiles and medians from data points up to 12\micron\ consistent with the right panel. In these objects, there is a more significant drop between the 4.6\micron\ data point and the 12\micron, but there is still observed flattening in the mid-IR.  

The 22\micron\ emission can also be produced by remnant hot dust emission from the recently quenched episode of star formation, which has been known to cause the overestimation of star formation in these types of sources \citep{hayward+14,utomo+14}. Sources that show these overestimates often show a decrease between the 12 and 22\micron\ bands, which is not the behavior that we observe in our sample in Figure~\ref{fig:psb_seds}. Alternative processes able to create a rising mid-IR SED at wavelengths shorter than 10\micron\ and provide sufficient emission to balance the stellar light from the galaxy are starbursts or AGNs. Starbursts have been ruled out by the weak nebular emission in the post-starburst systems \citep{goto07}, especially since the majority of post-merger star formation activity is expected to occur in the nucleus \citep{mihos+96}, at the position of the SDSS spectral fiber. Figure~\ref{fig:midIR_key} shows the 2--30\micron\ ranges of five phenomena that contribute to the mid-IR SED including stellar emission \citep{bruzual+charlot03}, AGB stars \citep{piovan+03}, star formation and PAH emission (NGC\,3521; \citealt{brown+14}),  a a starburst galaxy (Mrk\,33; \citealt{brown+14}), hot dust from an AGN \citep{sajina+12}. Both the starburst and AGN hot dust templates exhibit rising mid-IR emission, but it is improbable that the majority of post-starburst galaxies contain buried H\,{\sc ii} regions from prolific star formation. This leaves open the possibility that the remnant hot dust emission seen is being heated by low-luminosity AGNs.

Assuming that 100\% of the derived hot dust luminosity in the post-starburst galaxies originates from an AGN, we estimated the Eddington ratios for each 2MASS XSC post-starburst, which is detailed in Appendix~\ref{app:lumedd}. The resultant Eddington ratios range between 10$^{-2}$--10$^{-4}$, with a  peak around 10$^{-3}$. This firmly places these possible post-starburst AGNs into the low-luminosity regime. These Eddington ratios are also strict upper limits, as  it is unlikely that in all sources, 100\% of the hot dust luminosity is due to an AGN (since intermediate-aged stars also produce hot dust emission in post-starbursts as well).

To determine the X-ray properties of the 2MASS XSC post-starbursts, we cross-matched them with the {\em Swift} 70-month BAT All-sky Hard X-ray Survey \citep{baumgartner+13} but were unable to match any objects. Given the low Eddington ratio upper limits that the post-starburst 2MASS XSC samples, this is unsurprising, (the flux limit of the {\em Swift} BAT catalog is 1.03\,$\times$\,10$^{11}$~erg~s$^{-1}$~cm$^{-2}$, thus finds mainly high luminosity AGNs).

We also cross-matched the 2MASS XSC post-starbursts with the {\em Chandra} archive. Four were targeted by {\em Chandra} (ObsIDs 10270--10273; PI Zabludoff), and all four have weak-to-moderate detections above background fluctuations in a 2\arcsec\ aperture at the center of the galaxy. We use {\sc webpimms} to calculate the associated observed 2--10\,keV luminosities ranging between 10$^{40}$ to 6\,$\times$\,10$^{40}$, assuming a power law with $\Gamma$=1.8. The corresponding Eddington ratios (assuming $L_X$\,$\approx$\,1/16\,$L_{\rm bol}$; \citealt{ho2008}) range between 10$^{-5}$--10$^{-4.5}$.\footnote{One of the four sources (that with the highest X-ray flux) did not have sufficiently accurate spectroscopy to measure $\sigma_\star$, thus we were unable to derive an accurate Eddington ratio.} Another four were serendipitously in the fields of view of other {\em Chandra} observations of comparable exposure times but were not detected to significance above the background. That X-rays were detected coincident with the nuclei of many of the post-starbursts that were observed does provide further evidence that post-starbursts may contain an AGN phase, but given the small number statistics that still exist for these objects, is not conclusive.

Our estimate of possible Eddington ratios of the post-starbursts, derived from the hot dust luminosity, combined with the X-ray results seems to indicate that the presence of these low-luminosity AGNs in post-starbursts is entirely feasible given the estimated energetics. These results also support the conclusion of \citet{depropris+14}, that AGNs in post-starburst galaxies are not radiatively significant.  The H$\alpha$ fluxes necessary to pass the \citet{goto07} criterion seems to rule out radiatively significant AGNs, and in the case of the 2MASS XSC post-starbursts, do not likely represent Eddington ratios above $\sim$10$^{-3}$. Burying the AGN under a reservoir of optically thick gas is able to significantly impact the H$\alpha$ emission that can be observed, while the mid-IR AGN light is able to escape.

The E+A criterion selects against the presence of strong AGNs, quasars, and shocked systems (as these objects all emit H$\alpha$ or [O\,{\sc ii}] emission). Thus the possibility of AGNs being the source for the mid-IR colors is not a certainty. But new studies have opened up the possibility that there may be buried low-luminosity AGNs in post-starburst galaxies. Recently, significant molecular gas reservoirs have been discovered in post-starburst galaxies \citep{french+15,rowlands+15}. These CO-rich post-starbursts also exhibit excess 22\micron\ emission compared to their CO(1--0) emission, diverging from the relation set by the star-forming galaxies \citep{a16_spogco}. Additionally, the optical spectra of post-starbursts (despite having weak nebular lines) have LINER-like line emission, consistent with what is observed in low-luminosity AGNs \citep{yan+06,yang+06}. LINER emission also often originates from other sources, such as aged stellar populations \citep{yan+06,sarzi+10} and shocks \citep{allen+08,rich+11,a16_sample}, so this LINER emission is also not confirmatory of AGNs. The post-starburst composite SEDs and {\em WISE} colors support the possibility that many post-starbursts contain buried AGNs, which is not contradicted by the ionized gas line emission properties, and may even be supported by the presence of gas (and therefore an obscuring column).

A broad statistical analysis on emission line galaxies from SDSS showed that the AGN fraction in disk-dominated star-forming galaxies is significantly underestimated \citep{trump+15}, their signals being overwhelmed by the ionized gas signatures associated with star formation. \citet{bitsakis+15,bitsakis+16} note that in compact group galaxies (a known rapidly evolving population), once star formation started shutting down, the fraction of AGN-hosting galaxies increases even as the AGN luminosity decreases. These trends suggest either that weak AGN were always present in the nuclei of these galaxies but was being out-shined by star formation or that weak AGN activity begins during the phase of star formation quenching; it is unclear whether this result is universal.  Our results fuel further discussion about whether the quenching of star formation reduces the mid-IR signal that overwhelmed the weak AGN \citep{trump+15} or whether AGN fueling is part of the transition process \citep{hopkins+08}. Studying whether the mid-IR slope changes as a function of the stellar population age in post-starbursts may be able to discriminate between these two scenarios, but is beyond the scope of this paper.

\subsection{Toward a comprehensive selection of transitioning galaxies}
\label{sec:implications}
Selecting transitioning galaxies has long been a challenging endeavor. Optical colors can be ambiguous \citep{schawinski+14}; ultraviolet colors tend to be ultra-sensitive to star formation activity down to 1\% mass fractions \citep{kaviraj+07,kaviraj+07b,choi+09}, creating a set of ``frosted'' early-type galaxy interlopers. Robust spectral classification can be expensive, requiring high signal-to-noise spectra to detect absorption against the stellar continuum, placing the detection and cataloging high-redshift quenched galaxies out of reach. \citet{wild+14} showed that ``super-colors'' could be used to identify post-starburst-like galaxies in the redshift range 0.9\,$<$\,{\em z}\,$<$\,1.2, but this method has not been useful at lower redshift, due to its dependence on the ultraviolet portion of the SED.

Our new work opens yet another door to finding quenching galaxies, using the mid-IR colors. An independent investigation by \citet{ko+16} found that stacked spectra of mid-IR excess galaxies (defined using [3.4]--[12] {\em WISE} colors) showed signs of an intermediate stellar population. Post-starbursts, the bonafide transitioning population, sit in a distinct phase space in the {\em WISE} [3.4]--[4.6] vs. [4.6]--[12] colors, which, in the era of the {\em James Webb Space Telescope}, we will be able to observe up to {\em z}\,$\approx$\,1. 

Our work has also introduced a new challenge to how we identify quenching galaxies. Given that some post-starburst sources show signs of the presence of an AGN, it is likely that we are missing a significant population of quenching galaxies simply because we are removing all galaxies with significant emission in either H$\alpha$ or [O\,{\sc ii}], which an AGN will excite. That AGNs are present in E+A galaxies {\em whose selection criteria directly select against them} tells us that the AGNs are a crucial ingredient to study when trying to understand the nature of quenching galaxies, and should not be excluded when attempting to create a complete picture of galaxy metamorphosis, even at {\em z}\,=\,0 (and especially at high redshift).

\section{Summary}
\label{sec:summary}
We have analyzed the mid-IR properties of a selection of post-starburst galaxies selected through the ``E+A'' criterion by \citet{goto07} from SDSS DR7 \citep{sdssdr7}. Of the original 564 post-starbursts, we were able to analyze the colors of 534 objects with robust detections from both SDSS and {\em WISE} 3.4, 4.6, and 12\micron. We further investigated post-starbursts detected in the 2MASS XSC, totaling 190 objects, of which 158 have robust 3.4\micron, 4.6\micron\ and 12\micron\ detections. 53 of the 2MASS XSC post-starbursts are robustly (S/N\,$>$\,3) detected in the {\em WISE} 22\micron\ band. Using these samples, we came to the following conclusions.

The 534 post-starburst galaxies studied have transitioning \hbox{\em u--r} and [4.6]--[12] colors, falling within the infrared transition zone discussed by \citet{a14_irtz}.

After correcting for redshift effects, the [3.4]--[4.6] vs. [4.6]--[12] colors of post-starburst galaxies stand out from the colors of both early-type and late-type galaxies, inhabiting the mid-IR twilight zone. This result shows that galaxies do not transition directly across the mid-IR color gap between the early-type and late-type population, requiring an additional source of mid-IR emission.

The SED of post-starburst galaxies requires the inclusion of either strong neutral (11.3\micron) PAH emission or a TP-AGB component (with circumstellar dust) to fit the 3--12\micron\ data. A TP-AGB component would also be consistent with the findings of \citet{yan+06}, which required this component to explain the ionized gas emission seen in post-starbursts.

We used {\sc magphys} to fit the SEDs of the XSC post-starbursts, extracting the stellar and hot dust luminosities. We find that post-starbursts have intermediate hot dust luminosities (compared to the stellar), and that the [4.6]--[12]\micron\ {\em WISE} colors are a good proxy for the ratio between $L_\star$ and $L_{\rm hot~dust}$.

The composite SEDs of our observed post-starbursts (with 22\micron\ emission detected with S/N\,$>$\,3) suggest that an AGN component is needed to account for the hot dust detected in the 22\micron\ {\em WISE} band, but we cannot rule out other possibilities. Stacking the 22\micron\ emission in non-detected post-starbursts was also consistent with the need for a hot dust component to explain a flat mid-IR composite SED. The upper limit to the Eddington ratios inferred from the hot dust luminosity range between 10$^{-4}$--10$^{-2}$, with an average of 10$^{-3}$. This suggests that while AGNs might be present, they are low-luminosity and not radiatively dominant in the system.

Identifying galaxies that are transitioning requires a multiwavelength approach, and a closer look at the mid-IR has revealed new and exciting results. {\em WISE} colors suggest a path forward to photometrically identifying galaxies that are transitioning. SEDs of post-starbursts that include 22\micron\ emission suggest the presence of AGNs may be important for some of them, despite their ionized gas selection biasing against the presence of AGNs. We suggest that neglecting to allow for the presence of AGNs when selecting transitioning galaxies may be presenting a biased picture of how metamorphosis takes place.

\acknowledgments K.A. thanks R. Peletier for useful discussions as this manuscript was being prepared, as well as the anonymous referee for excellent suggestions that strengthened the manuscript. Support for K.A. is provided by NASA through Hubble Fellowship grant \hbox{\#HST-HF2-51352.001} awarded by the Space Telescope Science Institute, which is operated by the Association of Universities for Research in Astronomy, Inc., for NASA, under contract NAS5-26555. P.N.A. is partially supported by funding through {\em Herschel}, a European Space Agency Cornerstone Mission with significant participation by NASA, through an award issued by JPL/Caltech. T.B. would like to acknowledge support from the CONACyT Research Fellowships program. S.L.C. was supported by ALMA-CONICYT program 31110020. J.~F.-B. acknowledges support from grant AYA2016-77237-C3-1-P from the Spanish Ministry of Economy and Competitiveness (MINECO). L.L. acknowledges support for this work provided by NASA through an award issued by JPL/Caltech. K.N. acknowledges support from NASA  through the {\em Spitzer} Space Telescope. L.J.K. and A.M.M. acknowledge the support of the Australian Research Council (ARC) through Discovery project DP130103925. L.C. received funding from the European Union Seventh Framework Programme (FP7/2007-2013) under grant agreement n 312725. Support for A.M.M. is provided by NASA through Hubble Fellowship grant \hbox{\#HST-HF2-51377} awarded by the Space Telescope Science Institute, which is operated by the Association of Universities for Research in Astronomy, Inc., for NASA, under contract NAS5-26555.

This publication makes use of data products from the Wide-field Infrared Survey Explorer, which is a joint project of the University of California, Los Angeles, and the Jet Propulsion Laboratory/California Institute of Technology, funded by the National Aeronautics and Space Administration. The National Radio Astronomy Observatory is a facility of the National Science Foundation operated under cooperative agreement by Associated Universities, Inc.\\
{\it Facilities:} \facility{Sloan},\facility{WISE}

\vspace{-3mm}
\bibliographystyle{mnras}
\bibliography{../../master}

\appendix
\renewcommand{\thefigure}{A\arabic{figure}}
\begin{figure*}
\centering
\includegraphics[width=\textwidth]{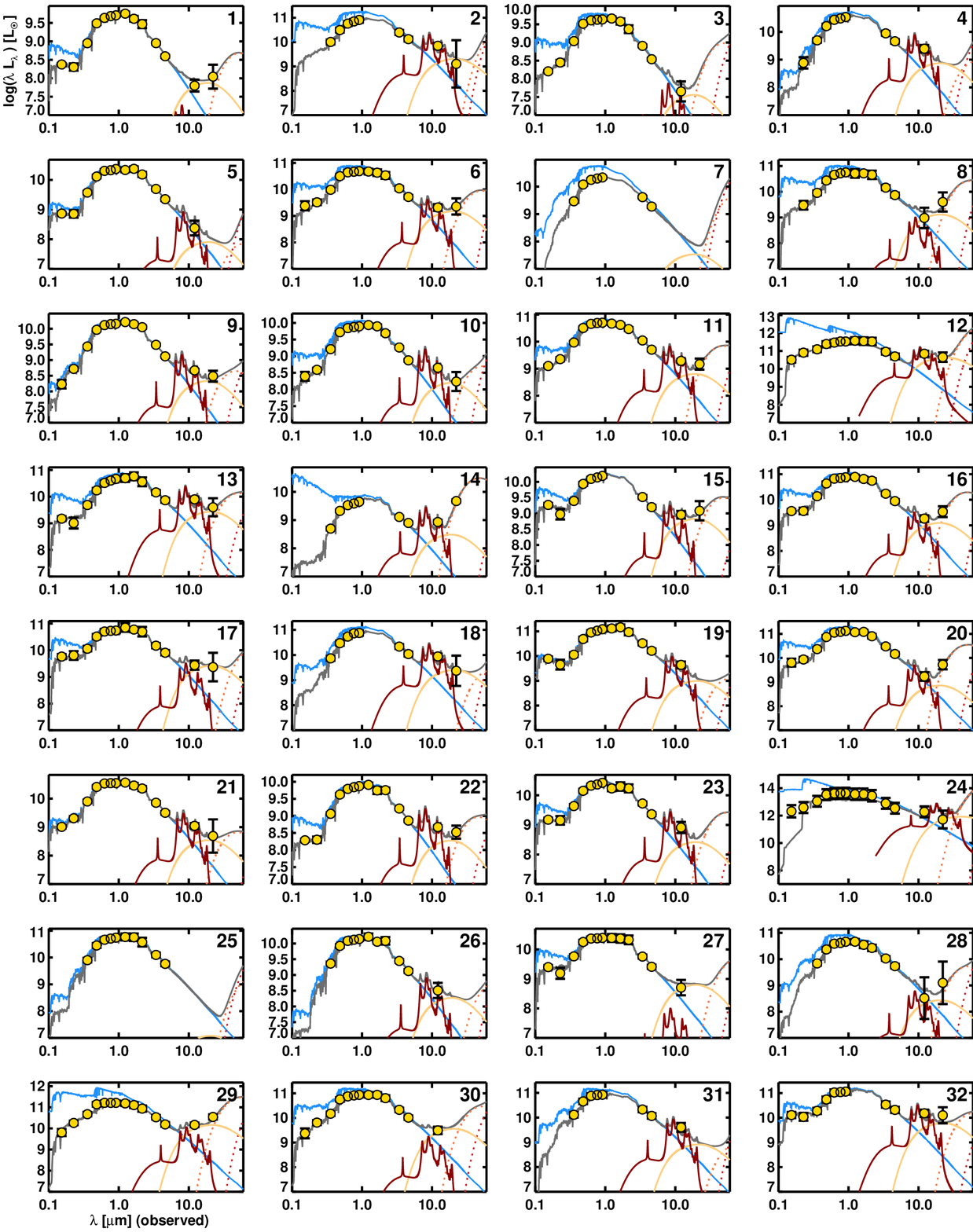}
\caption{{\sc magphys} models of the extended post-starburst galaxies. The yellow points are the observed photometric measurements for each post-starburst. The (unattenuated) stellar model is shown as a blue line. The hot (peach), warm (dotted salmon), and cold (dotted red) power laws are shown, as well as the expected PAH (maroon) emission. The complete fitted model is the dark gray line.}
\label{fig:magphys}
\end{figure*}
\renewcommand{\thefigure}{C\arabic{figure}}


\section{Fitting the Spectral Energy Distributions}
\label{app:magphys}
We performed UV to mid-IR SED fitting to estimate the physical properties of the galaxies both in our and the comparison samples. The code we used is called {\sc magphys} and it is described analytically in \citet{magphys}. It is based on the global energy balance between the energy absorbed in the UV and re-emitted in the IR, and adopts a Bayesian approach which draws from a large library of random models encompassing many parameter combinations, such as the star formation histories, metallicities and dust properties. The theoretical stellar models are computed by the \citet{bruzual+charlot03} population synthesis code, using the initial mass function presented in \citet{chabrier03}, whereas the dust models are from \citet{charlot+00}. The code compares the theoretical models with the observed SED of each galaxy and computes the $\chi^2$ value in order to build a probability distribution function (PDF) of each parameter. The final value of each parameter is thus the mean value of the PDF and the uncertainty associated is the given by the 16$^{\rm th}$ and 84$^{\rm th}$ percentiles of the distribution.

 {\sc magphys} fits multiple components to the re-emitted IR of each galaxy's SED, including PAH emission, two mid-IR dust components: hot ($T_{\rm dust}$\,=\,130 or 250\,K components) and one warm ($T_{\rm dust}$\,=\,30--60\,K) component, and finally a cold dust component. Figure~\ref{fig:magphys} shows the SEDs and {\sc magphys} fits to all 2MASS XSC post-starbursts. In the majority of cases, the {\sc magphys} fits are good, and are better in cases where both 12 and 22\micron\ emission is detected. The stellar (blue), PAH (maroon), and hot (peach) dust luminosities combine for the composite model (gray), which are shown in the panels in Figure~\ref{fig:magphys}.

\section{Comparison Sample Selection}
\label{app:compsample}
The comparison samples for the spirals and early-type galaxies used originate from the Galaxy Zoo project \citep{lintott+08,lintott+11}. The AGN sub-selection within Galaxy Zoo originates from \citet{schawinski+10}. The objects directly modeled were selected randomly from the corresponding catalogs. All objects are required to have available SDSS spectroscopic redshift and photometry in SDSS Data Release 9 \citep{sdssdr7}. 

The early-type galaxy and spiral samples were chosen from the Galaxy Zoo Data Release 2 catalog \citep{willett+13} having the corresponding morphological classifications (not being classified as AGN-hosting). They contain 232 and 194 galaxies respectively. 

The AGN sample was chosen from the AGN Host Galaxies catalog \citep{schawinski+10} and comprises 392 AGN-hosting galaxies. This catalog contains a volume-limited sample (0.02\,$<$\,$z$\,$<$\,0.05, $M_z$\,$<$\,-19.5\,AB) with emission line classifications consistent to those of AGN hosting galaxies. 

The UV data were obtained from the Galaxy Evolution Explorer All Sky Survey Data Release 6 \citep{morrissey+07}, resulting in far-ultraviolet (FUV; 1540\AA) and near-ultraviolet (NUV; 2300\AA) measurements. All photometric measurements were automatically performed using {\sc sextractor} (see \citealt{sextractor}). Finally, we used the {\em WISE} mid-IR photometry from \citet{lang+16}. These authors performed the ``forced photometry" technique in a consistent set of sources between SDSS and {\em WISE}, taking advantage of the high resolution of SDSS images to interpret the {\em WISE} data.

\section{Calculating the possible Eddington ratios from the hot dust luminosities}
\label{app:lumedd}
\begin{figure}[b!]
\vskip 2mm
\raggedright
\includegraphics[width=0.485\textwidth]{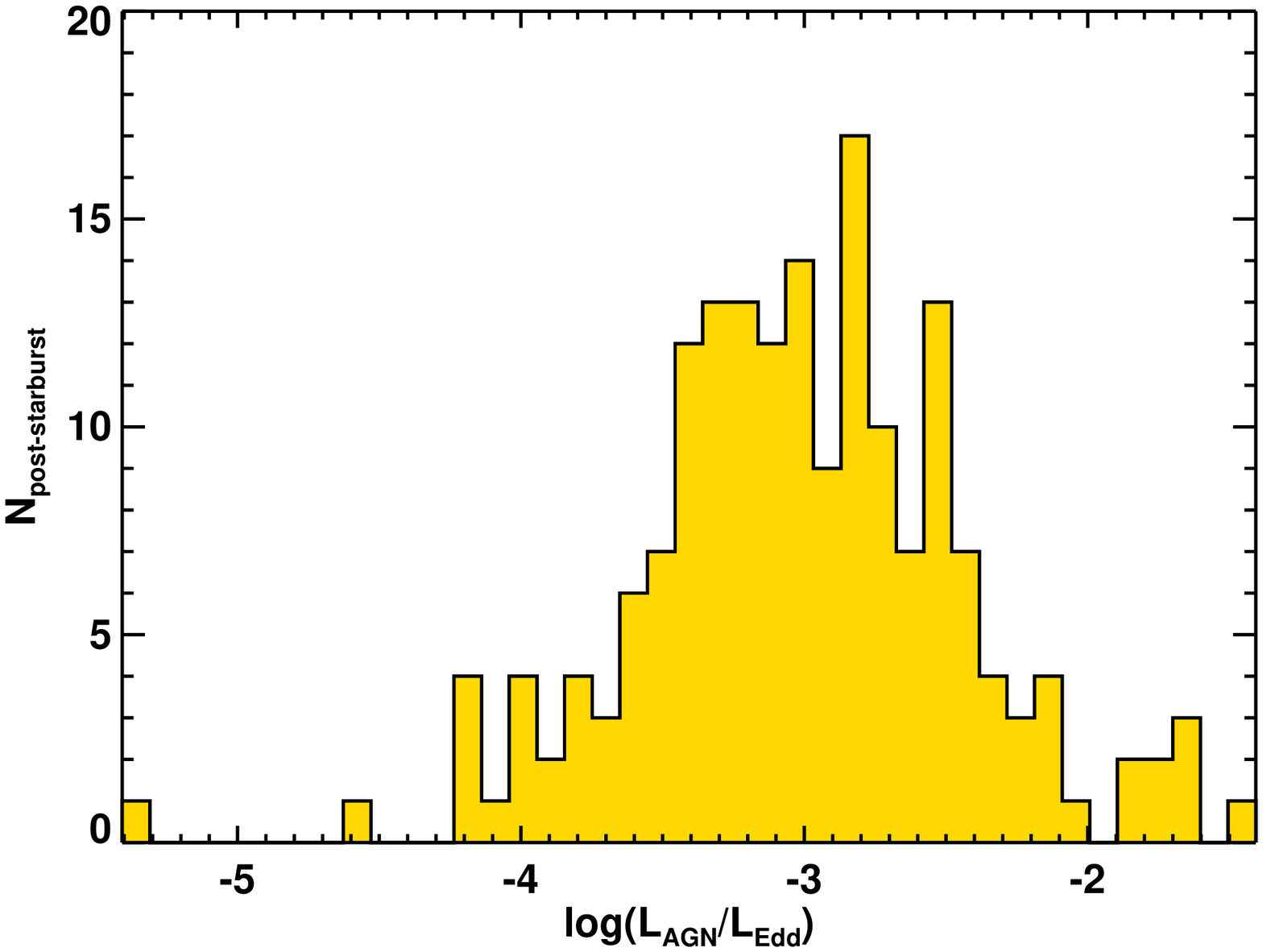}
\caption{The estimated upper limits to the Eddington ratios represented by the mid-IR hot dust luminosities of the post-starbursts. The estimated ratios firmly place the potential AGNs in post-starbursts into the low-luminosity AGN regime.}
\label{fig:eddratios}
\end{figure}

To determine the possible Eddington ratios of the post-starbursts, we derived values for black hole masses and potential AGN luminosities. We first obtained estimates of the stellar velocity dispersions from SDSS~DR9 \citep{sdssdr9} for the 190 2MASS XSC post-starbursts. We then estimated black hole masses using the $M_\bullet$-$\sigma$ relation from formula~7 in \citet{kormendy+13}. The Eddington luminosity was then derived using $L_{\rm Edd}/L_\odot$ = 3.2\,$\times$\,10$^4$~$M_\bullet$/$M_\odot$. The possible bolometric luminosities represented by these post-starbursts were estimated by assuming that 100\% of the {\sc magphys}-derived hot dust luminosity originated from an AGN \citep{sajina+05,richards+06,lusso+13}, thus we can estimate the bolometric luminosity of the AGN to be approximately 5$\times$\,$L_{\rm hot}$. 

Figure~\ref{fig:eddratios} shows the Eddington ratio distribution for the post-starbursts, assuming that the entire hot dust luminosity originates from an AGN. The derived Eddington ratios range between 10$^{-4}$ and 10$^{-2}$, with a peak around 10$^{-3}$. This firmly places the post-starbursts into the low-luminosity regime, consistent with the findings of \citet{brown+09} and \citet{depropris+14}, based on X-ray measurements.

\end{document}